\documentclass{article}

\usepackage[preprint]{neurips_data_2024}

\usepackage[utf8]{inputenc} 
\usepackage[T1]{fontenc}    
\usepackage{hyperref}       
\usepackage{url}            
\usepackage{booktabs}       
\usepackage{amsfonts}       
\usepackage{nicefrac}       
\usepackage{microtype}      
\usepackage{xcolor}         
\usepackage{multirow}
\usepackage{graphicx} 
\usepackage{subcaption}
\usepackage{adjustbox}

\title{SIU: A Million-Scale Structural Small Molecule-Protein Interaction Dataset for Unbiased Bioactivity Prediction}

\author{
Yanwen Huang\textsuperscript{1}\thanks{Equal contirbution}\;\,\thanks{Work was done while Yanwen Huang was an intern at AIR.}\;\,,
Bowen Gao\textsuperscript{2}\footnotemark[1]\;\,, 
Yinjun Jia\textsuperscript{3}, 
Hongbo Ma\textsuperscript{4}, \\
\textbf{Wei-Ying Ma\textsuperscript{2}},
\textbf{Ya-Qin Zhang\textsuperscript{2}}, 
\textbf{Yanyan Lan\textsuperscript{2}}\thanks{Correspondence to \texttt{lanyanyan@air.tsinghua.edu.cn}}  \\
\\
\textsuperscript{1}Department of Pharmaceutical Science, Peking University \\
\textsuperscript{2}Institute for AI Industry Research (AIR), Tsinghua University \\
\textsuperscript{3}School of Life Sciences, Tsinghua University \\
\textsuperscript{4}Department of Computer Science and Technology, Tsinghua University \\
}

\begin{document}

\maketitle

\begin{abstract}

Small molecules play a pivotal role in modern medicine, and scrutinizing their interactions with protein targets is essential for the discovery and development of novel, life-saving therapeutics. The term "bioactivity" encompasses various biological effects resulting from these interactions, including both binding and functional responses. The magnitude of bioactivity dictates the therapeutic or toxic pharmacological outcomes of small molecules, rendering accurate bioactivity prediction crucial for the development of safe and effective drugs. However, existing structural datasets of small molecule-protein interactions are often limited in scale and lack systematically organized bioactivity labels, thereby impeding our understanding of these interactions and precise bioactivity prediction. In this study, we introduce a comprehensive dataset of small molecule-protein interactions, consisting of over a million binding structures, each annotated with real biological activity labels. This dataset is designed to facilitate unbiased bioactivity prediction. We evaluated several classical models on this dataset, and the results demonstrate that the task of unbiased bioactivity prediction is challenging yet essential. 

\end{abstract}

\section{Introduction}

Bioactivity encapsulates various types of distinct measurements derived from different wet lab conditions, including both binding affinity and the spectrum of biological effects resulting from small molecule-protein interactions. Accurate bioactivity prediction is fundamental for discerning therapeutic potential and off-target toxicity effects, guiding medicinal chemistry efforts in discovering and optimizing potential small-molecule therapeutics, and is thus pivotal to the development of safe and effective drugs \cite{tropsha2024integrating, gaulton2010role}. For a small molecule to modulate the function of its protein target and exert its biological effects, it must be recognized by the protein through three-dimensional complementarity of shape and properties \cite{verma20103d}. This level of detail is indispensable, as knowing only the materials of a lock (protein) and the blueprint of a key (small molecule) is insufficient; 3D information is required to understand how the key fits into the lock and functions \cite{koshland1995key, eschenmoser1995one}. Current challenges in bioactivity prediction largely stem from the scarcity of high-quality, 3D structural data on small molecule-protein interactions.

The existing structural data on small molecule-protein interactions are markedly insufficient. Structural data derived from wet-lab experiments are limited, owing to the laborious and time-consuming nature of these assays. Additionally, this data often lacks comprehensive bioactivity annotations and is poorly organized with respect to bioactivity assay types. While computational modeling approaches have been employed to generate structural datasets, these efforts have yielded datasets of modest size with limited molecular diversity. The paucity of high-quality structural data still imposes a significant barrier to accurate bioactivity prediction, highlighting the critical need for computationally generated, large-scale, high-quality datasets. 

 To address this critical need, we present \textbf{SIU}: a million-scale \textbf{S}tructural small molecule-protein \textbf{I}nteraction dataset for \textbf{U}nbiased bioactivity prediction,\textbf{ the largest and most comprehensive structural dataset available to date}. SIU comprises over 5.34 million conformations, integrating both structural and bioactivity information for small molecule-protein interactions. Within this dataset, small molecule-protein pairs feature 1.38 million rigorously curated bioactivity annotations, each with a clear assay type designation. Our dataset provides extensive coverage of diverse small molecules, encompassing both active and inactive compounds, thereby surpassing the limitations of datasets restricted to molecules structurally similar to co-crystal ligands. It also includes a wide array of protein targets, covering all common protein classes, with each protein associated with multiple PDB IDs reflecting distinct pocket conformations. Our robust data generation pipeline employs multi-software docking and consensus filtering approach to ensure the precise modeling of small molecule-protein complexes. Bioactivity labels are meticulously curated and systematically organized according to assay types. SIU represents a significant advancement, offering a solid foundation for unbiased bioactivity prediction and enabling more accurate and comprehensive Pharmaceutical investigations.

We conducted experiments with several classical baseline models, and the results demonstrate that \textbf{our large-scale dataset can improve model performance compared to the widely used PDBbind dataset \cite{wang2004pdbbind, wang2005pdbbind}.} Additionally, the correlation results calculated by mixing different protein-molecule pairs are significantly higher than the correlation calculated after grouping by PDB IDs and using molecules for a single protein pocket. This indicates that \textbf{the correlation within PDB IDs is more challenging and serves as a more important metric for evaluating the bioactivity prediction ability of models.}s It highlights the importance of the unbiased bioactivity prediction task we introduced. This task focuses on the bioactivity difference for different molecules within a protein pocket, instead of the bias introduced by the different bioactivity ranges for different protein pockets.

In conclusion,\textbf{ our main contributions are threefold:}\textbf{ (1) }We introduced a million-scale structural dataset to address the exigent demands of the AI-driven drug discovery (AIDD) community;\textbf{ (2) }We devised and rigorously validated a robust, scalable pipeline for producing high-fidelity structural data of small molecule-protein interactions;\textbf{ (3) }We accentuated the significance of differentiating among bioactivity assay types and meticulously curated the dataset to enhance this practice in the training and evaluation of bioactivity prediction models.














\section{Related work}

\paragraph{Non-structural datasets on drug-target interaction for bioactivity prediction.}

A multitude of datasets are available for drug-target affinity (DTA) prediction; however, these datasets frequently lack structural data concerning the interactions between small molecules and their corresponding targets \cite{ekins2017chapter}.
 Large-scale bioactivity databases such as ChEMBL \cite{mendez2019chembl, gaulton2012chembl}, PubChem \cite{kim2016pubchem}, GuideToPharmacology \cite{pawson2014iuphar}, and DrugBank \cite{wishart2018drugbank,law2014drugbank} are invaluable resources. 
 Research efforts in DTA prediction primarily focus on binding affinity labels, with datasets such as Davis \cite{davis2011comprehensive} and  KIBA \cite{tang2014making} being widely utilized.
 MoleculeNet \cite{wu2018moleculenet} also comprises non-structural bioactivity data.

\paragraph{Structural datasets based on experimental structures for bioactivity prediction.}

The Protein Data Bank (PDB) \cite{berman2000protein}, established in 1971, has been an indispensable resource for structural biology, providing extensive structural data on protein and other biomolecules. However, it lacks direct bioactivity annotations and a systematic categorization of small molecules, often including non-specific and biologically irrelevant compounds. To address these limitations, several specialized databases have been developed, including PDBbind \cite{wang2004pdbbind, wang2005pdbbind}, Binding MOAD \cite{hu2005bindingmoad}, KiBank \cite{zhang2004development}, AffinDB \cite{block2006affindb}, and BioLiP \cite{yang2012biolip, wei2024q}. 
Their curation significantly enhances the utility of these datasets for structure-based Pharmaceutical research. Nevertheless, the reliance on labor-intensive experimental data acquisition limits the scalability and rapid expansion of these databases. Moreover, some databases lack explicit guidelines for the non-mixed use of assay types.

\paragraph{Structural datasets based on modeling structures for bioactivity prediction.}

Computational methods have been employed to construct datasets modeling small molecule-protein interaction structures that correspond with experimental bioactivities. The Natural Ligand DataBase (NLDB) \cite{murakami2016nldb} includes 7,053 complex structures, some of which are computationally modeled. The eModel-BDB \cite{naderi2018model} reports 200,005 structural entries, though it encounters issues such as steric clashes \cite{li2024high}. BindingNet \cite{li2024high} represents a novel dataset comprising 69,816 high-quality modeled structures obtained through comparative complex structure modeling. This modeling technique, however, requires the modeled small molecules to have structural similarity to co-crystal ligands from experiments, thereby limiting the quantity and diversity of the small molecules modeled in BindingNet.

\section{SIU dataset construction and overview}

The SIU dataset is a pioneering resource for predicting bioactivity, offering a comprehensive collection of small molecule-protein interactions with meticulously annotated bioactivity information. This section details the construction methods employed to ensure the data's quality, diversity, and organization for downstream tasks.

\begin{figure}[h]
    \centering
    \includegraphics[width=\textwidth]{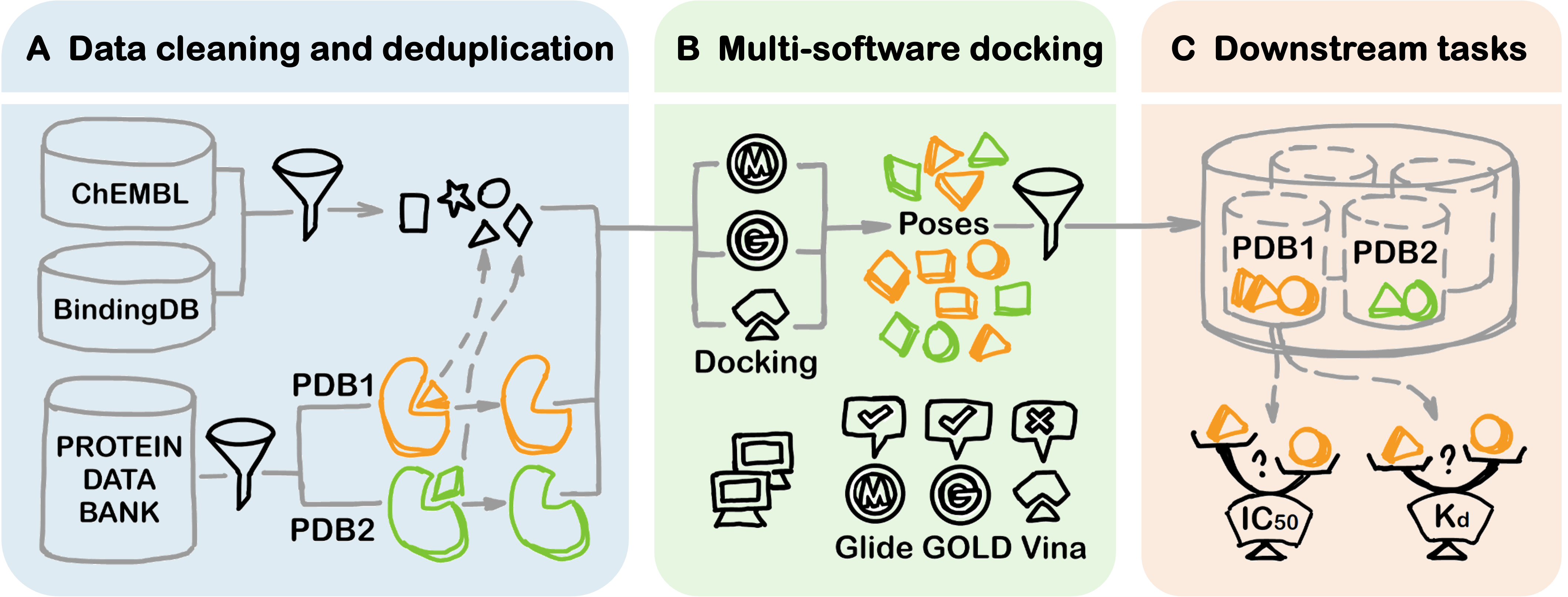}
    \caption{\textbf{Pipeline for SIU construction.} \textbf{(A)} Small molecules and protein targets were obtained from corresponding databases, cleaned, and deduplicated. Different small molecules binding to the same protein and different pockets (different PDB IDs) of the same protein were filtered and analyzed. \textbf{(B)} These were then subjected to a multi-software docking pipeline, where the small molecules were prepared and docked to their wet-experiment confirmed targets using three different software programs. The resulting poses were filtered through a voting mechanism to construct the final dataset. \textbf{(C)} The dataset is well-organized and contains multiple pockets for each protein and multiple molecules for each pocket, allowing for downstream tasks to be performed PDB-wisely and assay-type-wisely.}
    \label{fig:figure1}
\end{figure}

\subsection{Methods}

\subsubsection{Data cleaning and deduplication}

\paragraph{Bioactivity data extracting.} We retrieved non-structural bioactivity data from established databases: ChEMBL \cite{mendez2019chembl, gaulton2012chembl} and BindingDB \cite{chen2001bindingdb, liu2007bindingdb, gilson2016bindingdb}. Molecules were filtered based on predefined criteria.
Assays measuring bioactivity of small molecule-protein interactions were selected and filtered. The protein target information for each assay was carefully identified and standardized using UniProt IDs \cite{uniprot2015uniprot, uniprot2017uniprot}, ensuring consistency across datasets and facilitating matching with experimental structural data.

The small molecule filtering criteria are well-defined to exclude molecules that are not drug-like, including molecular weight, atom composition, and element restrictions. 
All small molecules retained their original IUPAC International Chemical Identifier (InChI) keys \cite{heller2015inchi} and Simplified Molecular Input Line Entry System (SMILES) notations \cite{weininger1988smiles, weininger1989smiles} from the databases to avoid mismatches due to different software calculations. Additionally, docking structurally similar small molecules for a single target leads to resource wastage. We examined targets associated with an excessively high number of small molecules and introduced a new filter based on small molecule extended-connectivity fingerprints (ECFP) similarity \cite{rogers2010extended},
ensuring both quality and diversity of small molecules while minimizing the computational expense of molecular docking.

The bioactivity data filtering process is also rigorously defined to ensure high-quality. Data from ChEMBL and BindingDB were independently extracted and cleaned before merging. For ChEMBL, criteria included assays involving only a single protein target, assays being either binding or functional, and bioactivity labels having standard relations, values, and units (i.e., $pM$, $nM$, or ${\mu}M$).
BindingDB data were extracted using similar logic, with slightly different filters due to database differences. The cleaned datasets were merged using InChI keys for small molecules and UniProt IDs for protein targets, ensuring precise matching of bioactivity labels to their respective small molecule-protein interactions. All small molecule-protein pairs with matched bioactivity labels were subsequently docked. The bioactivity information was standardized to a unit of $mol/L$ ($M$) and anti-logged, similar to datasets for drug-target binding affinity prediction \cite{ozturk2018deepdta}.

\paragraph{PDB structure retrieval and mapping.}

The protein structures were downloaded and matched with their UniProt IDs to ensure accurate alignment with bioactivity data. These structures were parsed into individual PDB format files, each representing a distinct pocket. Identified by PDB IDs, pockets are functional regions of the protein that interact with small molecules. We developed a filtering mechanism that leverages chemical and biological knowledge to eliminate PDB files containing non-specific or biologically irrelevant co-crystallized ligands not occupying genuine binding sites. The number of PDB files associated with each protein target varied significantly.
Docking all these structures is computationally expensive and offers diminishing returns in terms of novel information. We addressed this issue by implementing Fast Local Alignment of Protein Pockets (FLAPP) \cite{sankar2022fast} and other methods to further deduplicate the pocket library. As illustrated in Figure~\ref{fig:figure2}(A), this step efficiently removed highly similar pockets, resulting in a more streamlined pocket collection for docking simulations.

\subsubsection{Structural data construction via multi-software docking}


\paragraph{Molecular docking.}

SIU employs multiple docking software programs \cite{friesner2004glide, verdonk2003improved, trott2010autodock}, reducing reliance on any individual docking software. Initial 3D conformations for the small molecules were generated. For molecules with chiral centers, different stereoisomers were explored and included. Ionization states of molecules at physiological pH were also considered to ensure accurate representations of their charged forms. Multiple conformations were prepared for each small molecule to account for their flexibility.
The preprocessed data were organized into formats compatible with the chosen docking software. The protein targets were prepared, and grid files were generated according to each software's specific requirements to ensure compatibility. Small molecules were then docked into the binding pockets of the protein structures.

\paragraph{Consensus filtering of docking poses.}

The molecular docking results undergo rigorous scrutiny to ensure the retention of only credible poses. SIU employs a stringent filtering process: only those docking poses that exhibit consistency across at least two out of three different docking software results are retained. This consensus-based approach mitigates the inclusion of erroneous or misleading docking poses, thereby augmenting the overall quality and reliability of the dataset.

\begin{figure}[h]
    \centering
    \includegraphics[width=\textwidth]{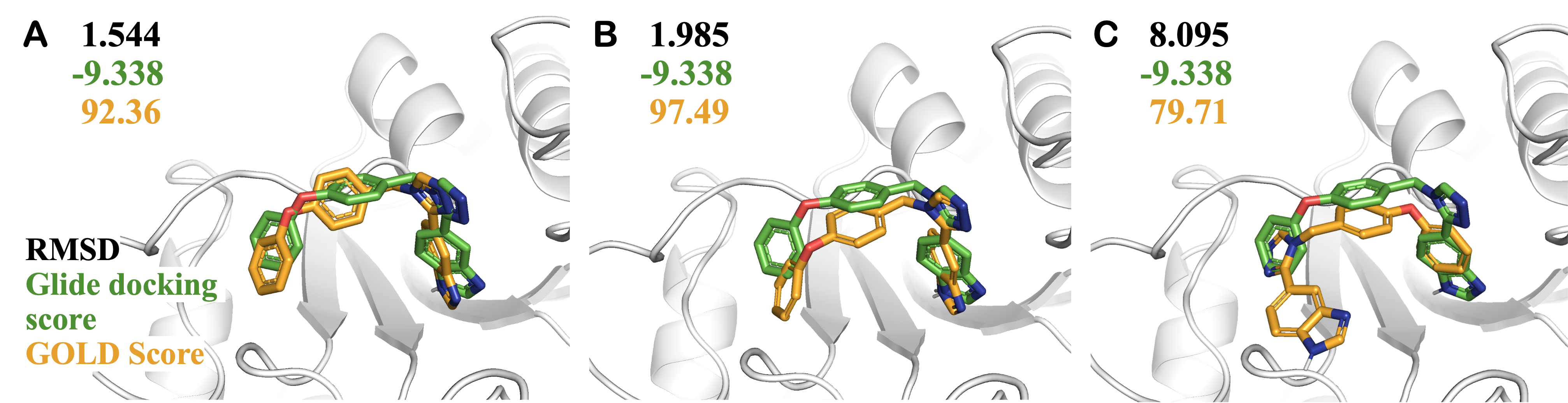}
    \caption{\textbf{Capability of RMSD to quantify differences in docking poses.} \textbf{(A)} RMSD 1.544, well-superimposed poses. \textbf{(B)} RMSD 1.985, similar binding modes. \textbf{(C)} RMSD 8.095, fundamentally different binding modes.}
    \label{fig:rmsd_vis}
\end{figure}

\begin{figure}[h]
    \centering
    \includegraphics[width=\textwidth]{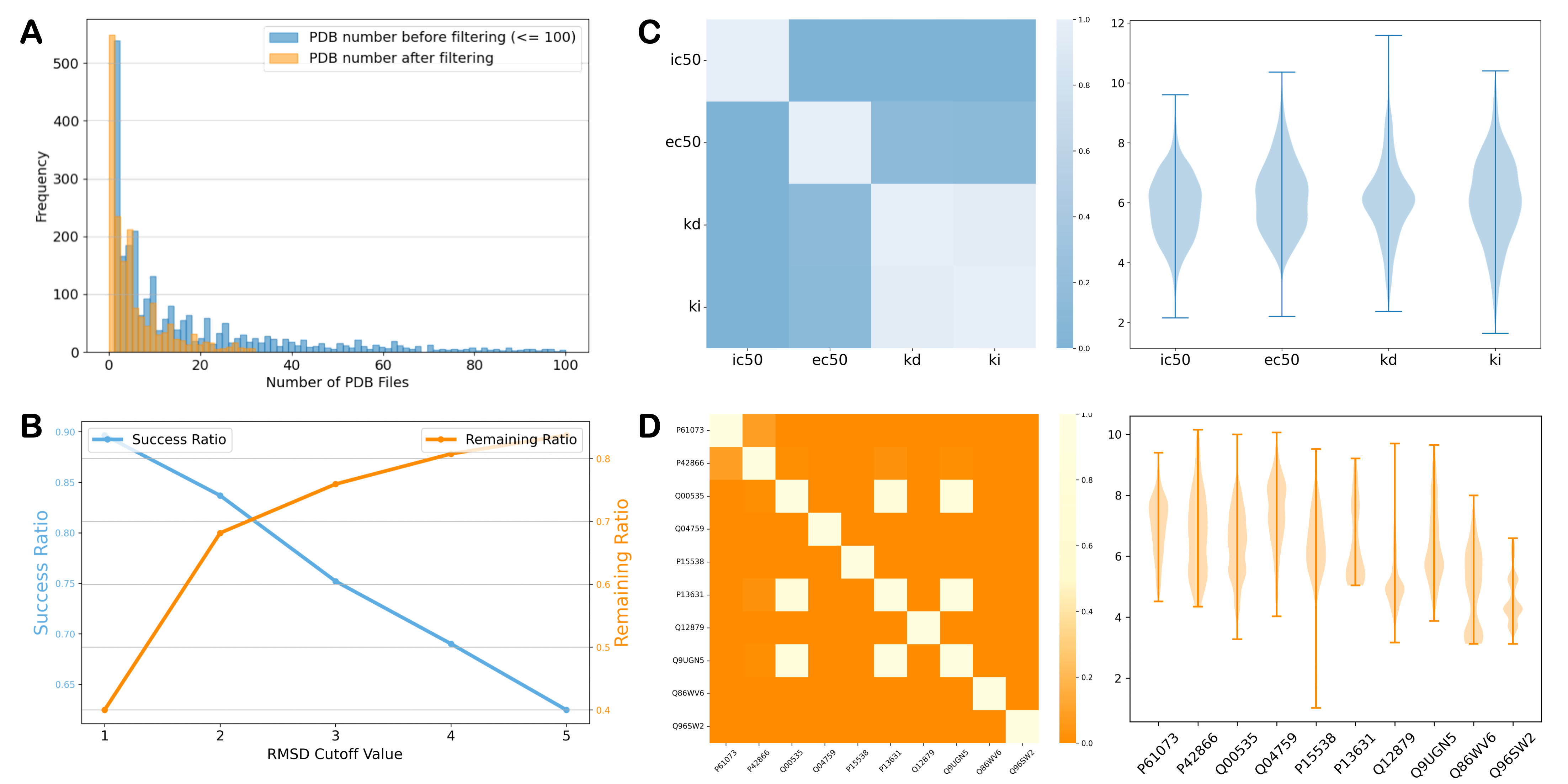}
    \caption{\textbf{Filter selection and dataset statistics.} \textbf{(A)} Distribution of the number of PDB files per protein target before and after filtering. \textbf{(B)} Influence of RMSD on success and retention ratios. \textbf{(C)} Pairwise t-test p-value differences between the negative logarithmic assay values of four representative assay types, visualized in a heatmap, along with the distribution of the values for each type. \textbf{(D)} Differences in assay values for ten representative protein targets, illustrated by a heatmap of their pairwise t-test p-values, and their distribution.}
    \label{fig:figure2}
\end{figure}

Different docking poses of the same small molecule-PDB pair were evaluated using the root mean square deviation (RMSD). Figure~\ref{fig:rmsd_vis} shows the RMSD and corresponding poses of a single Glide docking pose compared with the top three docking poses generated by GOLD.
When RMSD is about 2, the key interactions are maintained, indicating a potentially valid docking result.
This example underscores the importance of RMSD as a metric for evaluating the consistency and reliability of docking poses, with higher RMSD values indicating divergent binding modes that may arise from incorrect small molecule-protein interaction mode predictions. We further investigated
the trade-off between pose accuracy and the quantity of retained data, conducting experiments to observe the impact of varying RMSD values on these factors. Co-crystal poses, used as the ground truth, were extracted from PDB complexes and redocked into the original PDB pockets according to our docking procedure. The results, shown in Figure~\ref{fig:figure2}(B), indicate that when the RMSD is less than 2, a significant number of molecules can be retained, and the success ratio of the poses is satisfactory; as the RMSD increases, the number of retained poses slightly rises, but the accuracy of these poses significantly decreases. Therefore, an RMSD of 2 was selected as the cutoff.

\subsubsection{Data construction for downstream tasks}

\paragraph{Dataset organization for unbiased bioactivity prediciton.}

We organized data PDB-wisely and assay-type-wisely to facilitate unbiased bioactivity prediction, addressing the common issue of mixing the dissociation constant ($K_{d}$) \cite{lineweaver1934determination} and the inhibition constant ($K_{i}$) \cite{yung1973relationship} data and neglecting the data of other bioactivities. This meticulous organization supports the evaluation of small molecule-protein interactions with high fidelity, ensuring that the inherent differences in PDB files and assay types are respected.


\paragraph{Dataset split.}

To ensure the generalizability of the experimental findings with SIU, we employed a manual curation approach for dataset splitting. We selected a set of 10 representative protein targets to serve as the test set. These targets were intentionally chosen to cover a diverse range of protein classes, including well-known drug targets such as G-Protein Coupled Receptors (GPCRs) \cite{hauser2017trends}, kinases \cite{attwood2021trends, cohen2021kinase}, and cytochromes \cite{danielson2002cytochrome}. This selection strategy was designed to encompass the bioactivity landscape across various protein functionalities, thereby enhancing the applicability of our results to a wider range of potential drug discovery applications. We conducted non-homology analyses at two levels, 0.6 and 0.9, to ensure the independence and diversity of the training and test sets. For both versions 0.9 and 0.6, we have 21528 data pairs allocated for testing. Specifically, version 0.9 includes 1250807 data pairs for training and validation, while version 0.6 includes 386,330 data pairs for these purposes.

\subsection{Dataset overview}

\paragraph{Large-scale.} 

The SIU dataset comprises 5,342,250 conformations detailing small molecule-protein interactions, each entry providing comprehensive structural and bioactivity information. It includes 1,385,201 bioactivity labels derived from wet experiments, each with standardized values and clear assay type annotations. 

\paragraph{Diversity.}

SIU offers an extensive range of data, encompassing 214,686 diverse small molecules and 1,720 distinct protein targets. It includes experimentally validated low-bioactivity or inactive molecules, often absent in structural datasets from wet experiments, thus providing valuable negative data for AIDD. The dataset features extensive protein pocket coverage, including protein from humans, \textit{E. coli} \cite{vila2016escherichia}, various viruses, and other organisms. It spans major protein classes such as GPCRs \cite{hauser2017trends}, kinases \cite{attwood2021trends, cohen2021kinase}, nuclear receptors \cite{robinson2003nuclear}, cytochromes \cite{danielson2002cytochrome}, ion channels \cite{ashcroft1999ion}, and other protein involved in complex biological processes like epigenetics \cite{gibney2010epigenetics, feinberg2008epigenetics} and transcription \cite{cramer2019organization, lambert2018human}. As illustrated in Figure~\ref{fig:figure2}(D), the assay values of different protein targets vary significantly. This broad coverage ensures a comprehensive representation of small molecule-protein interaction modes, enhancing the relevance of our bioactivity prediction tasks to real biological environments.

\paragraph{High-quality.}

The structural information on small molecule-protein interactions in SIU is of high quality, due to our multi-software voting mechanism that maximizes docking accuracy within computational limits. As detailed in the structural data construction section, we achieved a satisfactory balance between data accuracy and scale, presenting high-quality data unobtainable with a single docking software or solely by ranking based on software-predicted docking scores. Docking software often provides successful simulated docking poses within the top-ranking positions, but these are not always ranked first by docking scores.
Our method, however, is based on the consistency of docking pose sampling across different algorithms. By examining consensus among different docking algorithms, we effectively ensure more accurate docking pose data.

\paragraph{Well-organized.}

SIU’s bioactivity labels are meticulously curated and systematically organized by PDB IDs and assay types, ensuring data integrity and enabling effective PDB-wise and assay-wise comparisons. This organization offers a robust resource for unbiased bioactivity prediction, addressing the limitations of existing datasets that often fail to distinguish clearly between different bioactivity assay types. Traditional measurements of correlations in bioactivity prediction tasks are often ineffective due to the lack of clarity in existing datasets. SIU can also address this problem, ensuring more precise and meaningful analyses. Our structured approach facilitates nuanced assessments, such as evaluating the impact of specific small molecule modifications on protein interactions or comparing the efficacy of different compounds within the same protein pocket context.

We argue that \textbf{the assay types should not be merged due to their distinct characteristics}. The heatmap in Figure~\ref{fig:figure2}(C) presents the results of pairwise t-tests for the four assay types, revealing that half maximal inhibitory concentration ($IC_{50}$) differs the most in mean value compared to the other assay types, followed by half maximal effective concentration ($EC_{50}$). In contrast, the means of $K_i$ and $K_d$ are relatively similar. However, even when the means are not significantly different, the assay types cannot be considered equivalent due to their distinct behaviors. 
As demonstrated in Figure~\ref{fig:figure2}(C), the bioactivity assay values vary significantly, with their negative logarithmic values ranging from 2 to 11 and exhibiting markedly different distributions.
The distribution of $K_d$ is particularly unique, as its upper values are substantially higher. The $K_d$ distribution is more peaked, indicating a narrow concentration around the central values despite its broad range.

Moreover, SIU provides multiple small molecule 3D poses for each protein pocket, allowing for unbiased comparison of small molecule poses while maintaining a constant protein pocket environment. This approach yields detailed information on how variations in small molecule poses influence their interactions with the protein pocket, considering factors such as shape and electrostatic complementarity. Ultimately, this enhances the modeling of the relationship between these interactions and observed bioactivity, advancing our understanding of small molecule-protein interactions and their effects.

\section{SIU experiments and analysis}
\begin{table}[ht]
\centering
\caption{Results for multi task learning with different label types. We show results for 3D-CNN, GNN, Uni-Mol, and ProFSA trained on SIU 0.9 version.}
\label{tab:mtl}
\begin{adjustbox}{width=0.9\textwidth}
\begin{tabular}{cc|cccccc}
\toprule
&   & RMSE & MAE & Pearson & Pearson\(^*\) & Spearman & Spearman\(^*\) \\ \hline
\noalign{\vskip 2pt}
\multirow{4}{*}{\begin{tabular}[c]{@{}c@{}}$IC_{50}$\end{tabular}} 
& 3D-CNN  & 1.560    & 1.275   & 0.158  & 0.044  & 0.154  & 0.040   \\
& GNN     &1.412     & 1.141   & 0.336  & 0.241  & 0.316  & 0.235   \\
& Uni-Mol & 1.353    & 1.092   & 0.462  & 0.343  & 0.466  & 0.351   \\
& ProFSA  & 1.361    & 1.108   & 0.382  & 0.331  & 0.356  & 0.317   \\ 
\noalign{\vskip 2pt}
\hline
\noalign{\vskip 2pt}
\multirow{4}{*}{\begin{tabular}[c]{@{}c@{}}$EC_{50}$\end{tabular}} 
& 3D-CNN  & 1.518    & 1.234   & 0.128  & 0.010  & 0.128  & 0.004   \\
& GNN     & 1.334    & 1.025   & 0.444  & 0.108  & 0.481  & 0.120   \\
& Uni-Mol & 1.273    & 1.017   & 0.428  & 0.178  & 0.461  & 0.144   \\
& ProFSA  & 1.255    & 0.971   & 0.438  & 0.204  & 0.495  & 0.154   \\ 
\noalign{\vskip 2pt}
\hline
\noalign{\vskip 2pt}
\multirow{4}{*}{\begin{tabular}[c]{@{}c@{}}$K_i$\end{tabular}} 
& 3D-CNN  & 1.534    & 1.260   & 0.201  & 0.025  & 0.200  & 0.021   \\
& GNN     & 1.814    & 1.504   & 0.247  & 0.099  & 0.107  & 0.058   \\
& Uni-Mol & 1.390    & 1.133   & 0.375  & 0.092  & 0.324  & 0.056   \\
& ProFSA  & 1.374    & 1.142   & 0.405  & 0.149  & 0.365  & 0.127   \\ 
\hline
\noalign{\vskip 2pt}
\multirow{4}{*}{\begin{tabular}[c]{@{}c@{}}$K_d$\end{tabular}} 
& 3D-CNN  & 1.503    & 1.233   & 0.173  & 0.024  & 0.167  & 0.038   \\
& GNN     & 1.711    & 1.431   & -0.068  & 0.065  & -0.147  & 0.033   \\
& Uni-Mol & 1.429    & 1.223   & -0.084  & 0.155  & -0.175  & 0.144   \\
& ProFSA  & 1.546    & 1.334   & -0.172  & 0.057  & -0.205  & 0.029   \\
\noalign{\vskip 2pt}
\bottomrule
\end{tabular}
\end{adjustbox}
\end{table}

We conducted experiments using several baseline models to analyze our SIU dataset. The models tested include a voxel-grid based 3D-CNN model, a Graph Neural Network (GNN) model, and pretrained models such as Uni-Mol \citep{zhou2022uni} and ProFSA \citep{gao2023self}. Our experiments were performed in both Multi-Task Learning (MTL) and single-target settings. In the MTL setting, all data were combined to train a single MTL model. In the single-target setting, the Uni-Mol model was trained separately on individual labels. 

The metrics used in our analysis include Root Mean Square Error (RMSE), Mean Absolute Error (MAE), general Pearson and Spearman correlation, and the correlation after grouping by PDB IDs. The general Pearson and Spearman correlations are calculated by mixing pairs of protein pockets and molecules. The grouped correlation metrics are calculated for different molecules within a single protein pocket. We use Pearson\(^*\) to represent Pearson correlation grouped by PDB IDs, and  Spearman\(^*\) to represent Spearman correlation grouped by PDB IDs.

Results for multi-task learning is shown in Table \ref{tab:mtl}, and the results for single task learning is shown in Talbe \ref{tab:single}.

\begin{table}[ht]
\centering
\caption{Results for single task training with different label types. We show the results with Uni-Mol model on PDBbind dataset, our SIU 0.6 version and 0.9 version dataset.}
\label{tab:single}
\begin{adjustbox}{width=0.9\textwidth}
\begin{tabular}{cc|cccccc}
\toprule
& Train Set  & RMSE & MAE & Pearson & Pearson\(^*\) & Spearman & Spearman\(^*\) \\ 
\hline
\noalign{\vskip 4pt}
\multirow{3}{*}{\begin{tabular}[c]{@{}c@{}}$IC_{50}$\end{tabular}} 
& PDBbind &  1.575   &1.279    &0.430   &0.245   &0.425   &0.229    \\
\noalign{\vskip 2pt}
& SIU 0.6 &1.407  &1.138    &0.461   &0.317   &0.463   &0.311   \\
\noalign{\vskip 2pt}
& SIU 0.9  &1.357  &1.099    &0.470   &0.345   &0.474   &0.347    \\ 
\noalign{\vskip 2pt}
\hline
\noalign{\vskip 2pt}
\multirow{2}{*}{\begin{tabular}[c]{@{}c@{}}$EC_{50}$\end{tabular}} 
& SIU 0.6 & 1.400    & 1.163   & 0.280  & 0.171  & 0.284  & 0.150   \\
& SIU 0.9  & 1.340    & 1.096   & 0.384  & 0.196  & 0.379  & 0.142   \\ 
\noalign{\vskip 2pt}
\hline
\noalign{\vskip 2pt}
\multirow{3}{*}{\begin{tabular}[c]{@{}c@{}}$K_i$\end{tabular}} 
& PDBbind &1.315    &1.085  &0.368   &0.040   &0.323   &0.026    \\
& SIU 0.6 & 1.255    & 1.034   & 0.472  & 0.106  & 0.452  & 0.112   \\
& SIU 0.9  & 1.235    & 1.017   & 0.485  & 0.036  & 0.452  & 0.041   \\ 
\noalign{\vskip 2pt}
\hline
\noalign{\vskip 2pt}
\multirow{3}{*}{\begin{tabular}[c]{@{}c@{}}$K_d$\end{tabular}} 
& PDBbind & 1.565    & 1.308   & 0.041  & 0.010  & 0.004  & 0.006  \\
& SIU 0.6 & 1.389    & 1.192  & -0.149  & 0.052  & -0.206  & 0.022   \\
& SIU 0.9  & 1.364    & 1.141   & -0.033  & 0.103  & -0.082  & 0.065  \\ 
\noalign{\vskip 2pt}
\bottomrule
\end{tabular}
\end{adjustbox}
\label{tab:table2}
\end{table}

\begin{figure}[ht!]
    \centering
    \begin{subfigure}[b]{0.49\textwidth}
        \includegraphics[width=\textwidth]{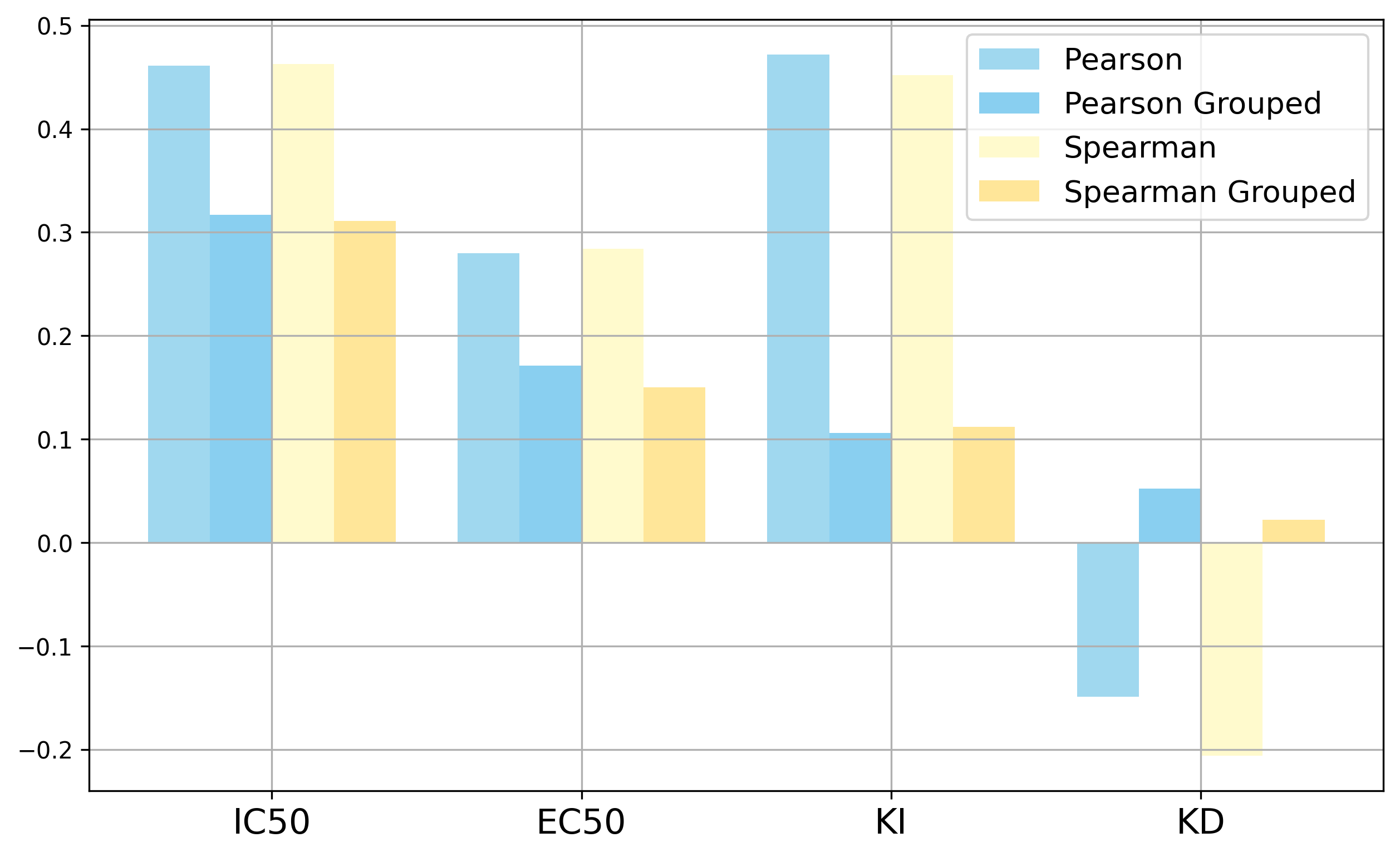}
        \caption{}
        \label{fig:cor agg}
    \end{subfigure}
    \hfill 
    \begin{subfigure}[b]{0.49\textwidth}
        \includegraphics[width=\textwidth]{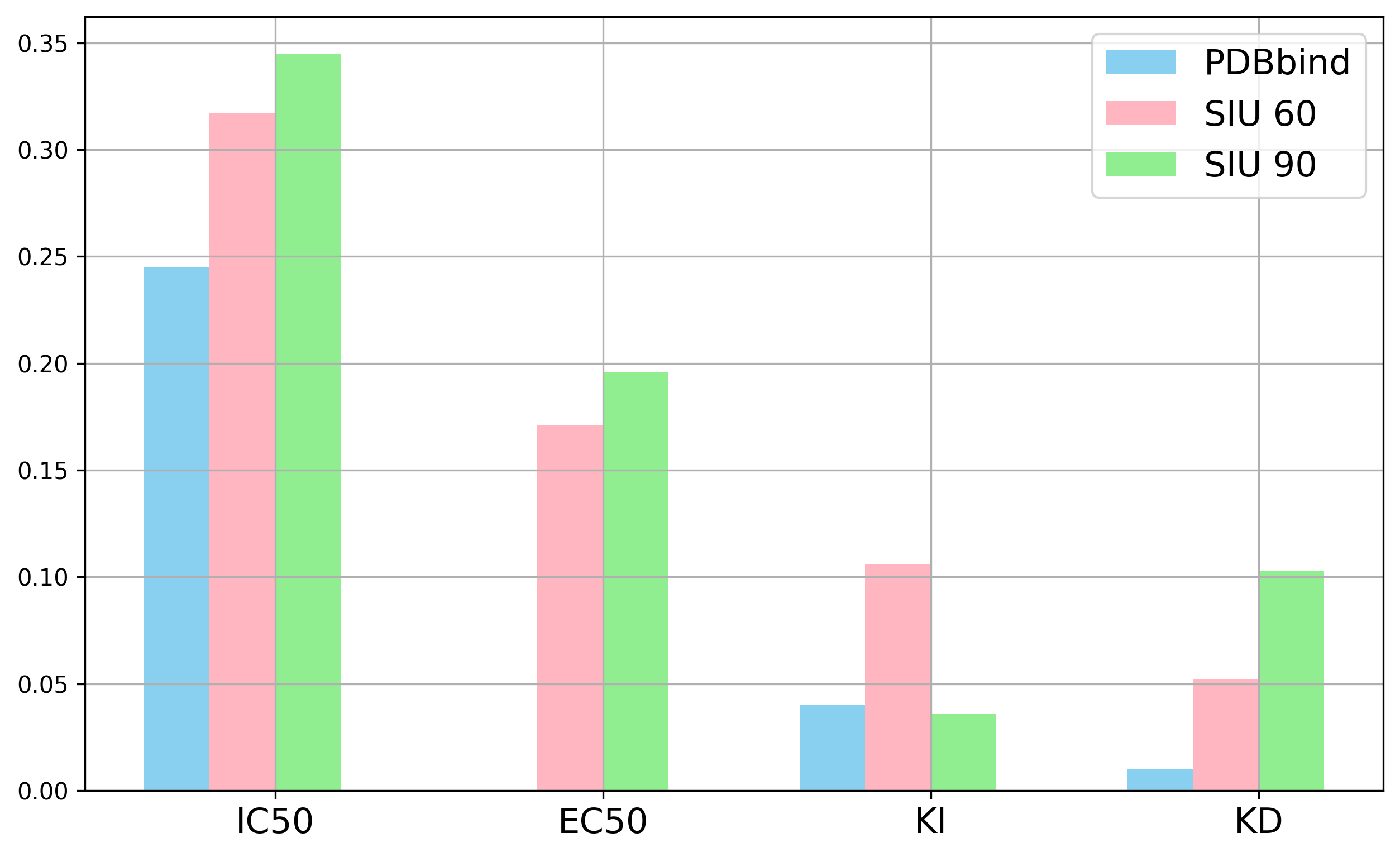}
        \caption{}
        \label{fig:sub2}
    \end{subfigure}
    \caption{(a) Pearson and Spearman correlations for various label types, calculated both before and after grouping by PDB IDs. (b) Pearson correlations after grouping PDB IDs for different assay types trained on different datasets.}
    \label{fig:cor dataset}
\end{figure}

\subsection{Analysis}

\paragraph{Different difficulties of assay types.}

The bioactivity prediction difficulty varies among different assay types. The $K_d$ task is the most challenging, primarily due to the varying correlations between different assay types, as shown in Figure~\ref{fig:figure2}(C). Although the means of $K_i$ and $K_d$ labels do not differ statistically, the correlation between these two data groups remains limited. The intrinsic differences in assay types of bioactivity arise from the principles of the wet-lab experiments used to measure them. Binding assays focus on the direct interaction between the small molecule and the protein target, providing insights into the strength and specificity of this binding through metrics like $K_i$ and $K_d$, using techniques such as surface plasmon resonance (SPR) \cite{schasfoort2017introduction, englebienne2003surface} and isothermal titration calorimetry (ITC) \cite{leavitt2001direct}. In contrast, functional assays measure the biological response elicited by the small molecule on the target, capturing its effect on a biological system and often quantified by $IC_{50}$ and $EC_{50}$ by enzyme activity assays \cite{bisswanger2014enzyme, hall1996methods} or other wet experiment techniques. \textbf{The inherent differences in what these assays measure mean that their values cannot be directly compared \cite{yung1973relationship}.} Furthermore, even within the categories of binding and functional assays, metrics should not be used interchangeably, as $K_i$ and $K_d$ describe different aspects of binding affinity, just as $IC_{50}$ and $EC_{50}$ describe different aspects of biological response.

\paragraph{Influence of measuring correlation with same PDB IDs.}

As demonstrated in Figure \ref{fig:cor agg}, aggregating data by pdb ID across all assay types results in a significant decline in both Pearson and Spearman correlations. This observation suggests that it is more challenging to achieve high correlation when assessing binding affinities for different molecules within the same pocket after grouping. This challenge primarily arises from the skewed distribution of binding affinities across various protein pockets, as shown in Figure~\ref{fig:figure2}(D). Furthermore, these findings highlight that \textbf{conventional approaches to measuring correlation without grouping by PDB ID may not effectively capture a model’s ability to differentiate between molecules targeting the same protein.} Such discriminatory capacity is crucial in drug discovery, emphasizing the importance of focusing on molecular interactions specific to each target rather than general correlations across diverse targets. This underscores the necessity of our dataset, which measures correlation within the same pdb IDs, providing a more relevant assessment of a deep learning model’s utility in drug discovery.

\paragraph{Effectivness of training on larger dataset.}

We compare models trained on the PDBbind 2020 dataset with those trained on SIU versions 0.6 and 0.9. Notably, the PDBbind 2020 dataset was used in its entirety, without implementing any filtering techniques to exclude pockets similar to those in the test set. As illustrated in Table~\ref{tab:table2} and Figure~\ref{fig:cor dataset}, models trained on the SIU datasets outperform those trained on PDBbind, despite the latter’s lack of homology removal. \textbf{This underscores the effectiveness of our large-scale dataset in enhancing model learning for binding affinity prediction.} Also the 0.9 version gives a better performance compared to the 0.6 version, indicating the influence of removing homology and scaling law of the dataset.

\section{Limitations and future work} \label{sec: limitation}

\paragraph{Limitations.}

Despite our rigorous methodology, the structural data we obtained are still predicted poses rather than experimentally validated interactions. The challenge of accurately modeling small molecule-protein interactions in physiological conditions remains substantial and highlights the need for continued advancements in this field.

\paragraph{Future work.}

We aim to provide larger and more reliable datasets for various drug discovery tasks. we are developing datasets for pairwise ranking, alongside organizing data for unbiased bioactivity prediction. We will ensure that comparisons are made only between docking poses derived from the same PDB and bioactivity data from identical assay types.
Additionally, our approach allowing the automated generation of extensive high-quality data has been validated as feasible and scalable in this work. Though in this work, to address the computational demands of the molecular docking stage, we optimized the dataset by deduplicating small molecules and pockect structures, Future research could build upon these methods to construct even larger datasets efficiently, thereby advancing the understanding of small molecule-protein interactions in AIDD.

\section{Conclusion}

To meet the pressing demands in drug discovery and development, we introduced SIU, a large-scale, diverse, accurate, and well-curated dataset for unbiased bioactivity prediction. SIU was constructed using meticulously designed and robust pipelines, ensuring its exceptional quality and comprehensiveness. Our experimental results validate the large scale nature of SIU as a superior training dataset that enhances model performance and provides a reliable framework for unbiased bioactivity prediction tasks, facilitating more meaningful model evaluations. We anticipate that the full potential of SIU remains to be uncovered, and we hope that its introduction will catalyze significant advancements in the field of drug discovery and development.

\bibliography{siu}
\bibliographystyle{plainnat}

\newpage

\appendix

\section{Dataset and Code Availability}

The whole dataset and corresponding descriptions can be found at \\ \href{https://huggingface.co/datasets/bgao95/SIU}{https://huggingface.co/datasets/bgao95/SIU}

The code and instructions used to train the baseline models can be found at \\ \href{https://github.com/bowen-gao/SIU}{https://github.com/bowen-gao/SIU}

The dataset is hosted by Hugging Face. The license is CC BY 4.0. We bear all responsibility in case of violation of rights.

The data we are using/curating doesn't contain personally identifiable
information or offensive content.

\section{Dataset overview}

SIU represents a large-scale, high-quality dataset of small molecule-protein interactions, meticulously organized to facilitate unbiased bioactivity prediction, both PDB-wise and assay-type-wise. The dataset comprises a total of 5,342,250 conformations. Each instance in the dataset provides detailed information about small molecule-protein interactions, including the coordinates and element types of each atom in the small molecule and the corresponding pockets of each interaction. Additionally, the assay value and type of each conformation, along with other critical information, are carefully obtained and retained from the original bioactivity databases. This includes the UniProt ID and PDB ID of the protein pockets, as well as the InChI keys \citep{heller2015inchi} and SMILES \cite{weininger1988smiles, weininger1989smiles} notations of the small molecules.

\begin{table}[ht]
\centering
\caption{The label count for 4 representative assay types in SIU total, SIU 0.9, and 0.6 versions.}
\label{tab:count}
\begin{tabular}{c|cccc|cccc}
\toprule

& \multicolumn{4}{c|}{\textbf{SIU 0.9 version}} & \multicolumn{4}{c}{\textbf{SIU 0.6 version}}\\
\cmidrule{2-5}
\cmidrule{6-9}
& Total & Train & Valid & Test & Total & Train & Valid & Test \\ 
\midrule
 $MTL$ & 1272335 & 1125727 & 125080 & 21528 & 407858 & 347697 & 38633 & 21528 \\
\noalign{\vskip 3pt}
 $IC_{50}$ & 962063 & 854230 & 94859 & 12974 & 320594 & 276969 & 30651 & 12974 \\

 $EC_{50}$ & 97952 & 84067 & 9508 & 4377 & 32842 & 25675 & 2790 & 4377 \\

 $K_i$ & 198091 & 175442 & 19447 & 3202 & 47946 & 40188 & 4556 & 3202 \\

 $K_d$ & 54570 & 47347 & 5347 & 1876 & 17509 & 14003 & 1630 & 1876 \\
\bottomrule
\end{tabular}
\end{table}

Additionally, the dataset encompasses over 1,385,201 million assay labels, each derived from corresponding wet-lab bioactivity experiments, ensuring the reliability and accuracy of the bioactivity information. SIU includes 1,720 diverse protein targets, with each protein potentially possessing multiple distinct binding pockets, verified through rigorous deduplication methods, resulting in a total of 9,662 unique pockets. The dataset also features a substantial and diverse collection of small molecules, totaling 214,686, across all pockets. Importantly, we have only included protein pocket-small molecule pairs confirmed to be active or inactive through wet-lab experiments, amounting to over 1,291,362 million pairs.

As shown in Table~\ref{tab:count}, we included various assay types with significant amounts of data to demonstrate that bioactivity values from different assay types should not be mixed and to facilitate the future assay-type-specific use of SIU. Additionally, to examine the structural differences among small molecules from the top four assay types, a random sample from each was visualized using t-SNE with ECFP fingerprints (radius = 3, 1024-bit vectors), as depicted in Figure~\ref{fig:figure1}.

\begin{figure}[h]
    \centering
    \includegraphics[width=0.7\textwidth]{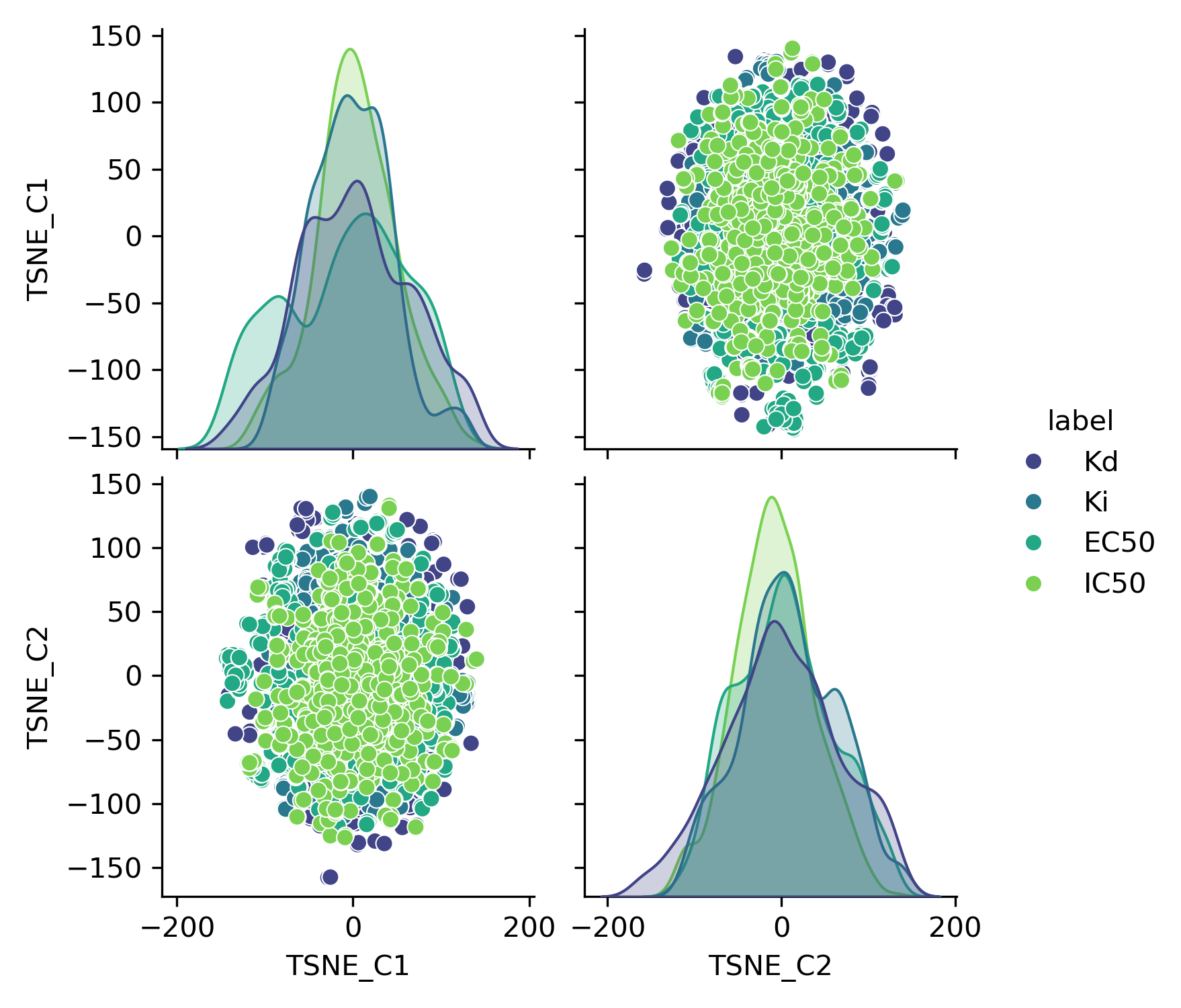}
    \caption{Visualization of chemical structure differences among small molecules from the top four assay types using t-SNE with ECFP fingerprints.}
    \label{fig:figure1}
\end{figure}

\section{Test set construction}

To ensure the robustness and generalizability of the experimental findings with SIU, we meticulously curated a test set composed of 10 protein targets, as listed in Table~\ref{tab:test}. These targets were selected to represent a wide range of protein classes, including G-Protein Coupled Receptors (GPCRs), kinases, cytochrome, nuclear receptor, ion channel, epigenetic, and others, ensuring broad coverage of the bioactivity landscape. For example, "C11B1\_HUMAN" belongs to the cytochrome P450 family, which is involved in the metabolism of various drugs \cite{bureik2002human, denisov2005structure}. "RARG\_HUMAN" belongs to the Nuclear Receptor family, with drugs like bexarotene used for certain cancers \cite{altucci2007rar, qu2010bexarotene}. "NMDE1\_HUMAN" represents the NMDA receptor, a critical glutamate receptor in neurons implicated in various neurological disorders, with memantine being an approved NMDA receptor antagonist for moderate to severe Alzheimer's disease \cite{mori1995structure, reisberg2003memantine}. Including these targets across various functionalities enhances the applicability of our results in drug discovery.

\begin{table}[h]
\centering
\caption{The curated test set of 10 protein targets, covering a diverse range of protein classes and displaying an even distribution of small molecule-pocket pair counts.}
\label{tab:test}
\begin{tabular}{c|c|c|c}

\toprule

UniProt & Gene name & Class & Small molecule- \\
 &  &  & pocket pair count\\
\midrule

P61073 & CXCR4\_HUMAN & GPCR & 1376 \\

P42866 & OPRM\_MOUSE & GPCR & 2379  \\

Q00535 & CDK5\_HUMAN & Kinase & 2189 \\

Q04759 & KPCT\_HUMAN & Kinase & 2320 \\

P15538 & C11B1\_HUMAN & Cytochrome & 2427 \\

P13631 & RARG\_HUMAN & Nuclear Receptor & 1888 \\

Q12879 & NMDE1\_HUMAN & Ion Channel & 2144 \\

Q9UGN5 & PARP2\_HUMAN & Epigenetic & 2251 \\

Q86WV6 & STING\_HUMAN & Others & 2495 \\

Q96SW2 & CRBN\_HUMAN & Others & 2059 \\

\bottomrule
\end{tabular}
\end{table}

\section{Model Training}

For GNN Model, we use the same model in atom3d \citep{townshend2021atom3d} \href{https://github.com/drorlab/atom3d/}{https://github.com/drorlab/atom3d/}. We train the model using one NVIDIA A100 GPU. The batch size is 256, the max number of epochs is 20, the optimizer is Adam, the learning rate is 1e-3.

For 3D-CNN Model, we use the same model in atom3d \citep{townshend2021atom3d} \href{https://github.com/drorlab/atom3d/}{https://github.com/drorlab/atom3d/}. We train the model using one NVIDIA A100 GPU. The batch size is 256, the max number of epochs is 20, the optimizer is Adam, the learning rate is 1e-4.

For Uni-Mol model, we use the pretrained model weights provided by \href{https://github.com/dptech-corp/Uni-Mol/}{https://github.com/dptech-corp/Uni-Mol/}. The pretrained molecular encoder and pocket encoder outputs are concatenated and passed through a four-layer Multi-Layer Perceptron (MLP) with hidden dimension 1024, 521, 256, 128. We use four NVIDIA A100 GPU to train the model. The batch size is 384, the max number of epochs is 50, the optimizer is Adam, the learning rate is 1e-4.

For ProFSA model, we use the pretrained model weights provided by \href{https://github.com/bowen-gao/ProFSA}{https://github.com/bowen-gao/ProFSA}. The pretrained molecular encoder and pocket encoder outputs are concatenated and passed through a four-layer Multi-Layer Perceptron (MLP) with hidden dimension 1024, 521, 256, 128. We use four NVIDIA A100 GPU to train the model. The batch size is 384, the max number of epochs is 50, the optimizer is Adam, the learning rate is 1e-4.

Details can found at  \href{https://github.com/bowen-gao/SIU}{https://github.com/bowen-gao/SIU}

\section{Potential negative impact of SIU}

While our dataset, SIU, represents a significant advancement in the field of bioactivity prediction, it is important to acknowledge potential limitations and areas of concern. Despite our robust multi-software docking approach and consensus filtering, the inherent reliance on computational methods may still introduce certain biases or inaccuracies in the modeled small molecule-protein interactions. These potential inaccuracies could inadvertently mislead researchers, leading to less reliable predictions and potentially diverting attention away from promising compounds or targets.

\end{document}